\documentclass[referee]{cjaa}
\usepackage{graphicx}

\begin{document}

\title{Pulsars in FIRST Observations}
\author{X.H. Li\mailto{}\and J.L. Han}
\offprints{X.H. Li}
\email{xhli@ns.bao.ac.cn}
\institute{ The Partner Group of MPIfR, National Astronomical Observatories, Chinese
Academy of Sciences, Beijing 100012}
\date{Received~~2001~~~~~~~~~~~~~~~; accepted~~2001~~~~~~~~~~~~~~~~}
\abstract{
We identified 16 pulsars from the Faint Images of the Radio Sky at
Twenty-cm (FIRST) at 1.4 GHz. Their positions and total flux densities are extracted
from the FIRST 
catalog. Comparing the source positions with those in the PSR
catalog, we obtained better determined positions of PSRs J1022+1001,
J1518+4904, J1652+2651, and proper motion
upper limits of another three pulsars PSRs J0751+1807, J1012+5307,
J1640+2224. Proper motions of the other 10 pulsars are
consistent with the values in the catalog.
}
\authorrunning{X.H. Li \& J.L. Han}
\maketitle
\section{INTRODUCTION}
Pulsars obtain a large kick velocity during their birth. Therefore pulsars
moved away from their birth place, which is believed to be the center of
supernova remnant. Measuring the proper motion of pulsars can derive an
independent measurment of pulsar age, which has a number of
astrophysical usages. For example, studies of the evolution of neutron star
magnetic fields and pulsar emission beam.

There are three methods to measure proper motions: (1) traditional optical
methods, (2) timing measurements, and (3) interferometric measurements. The
first technique is applied to a few pulsars which can be seen in optical
wavelengh (Mignani et al. 2000). Pulsar proper
motions can be obtained from regular timing observations carried out over a
sufficiently long 
interval. However, timing noise restricts its usefulness 
except millisecond pulsars (Kaspi, Taylor \& Ryba
1994; Bell et al.1995; Nice \& Taylor 1995; Wolszczan et al. 2000). Interferometric observations 
measure the angular transverse motions of the pulsars in the sky, relative to
a set of reference sources (e.g. McGary et al. 2001). This technique
is the most
productive methods for determing proper motions of pulsars so far.

FIRST (Becker et al. 1995) is a project designed to produce the radio equivalent of the Palomar
Observatory Sky Survey over 
10,000 square degrees of the North Galactic Cap ----- using B-configuration of
NRAO Very Large Array (VLA) at 1.4 GHz. The observations have a
resolution of $5^{''}$, and the positional accuracies have 90\% confidence
error circles of radius $<$0.5$''$ at the 3 mJy level and $1^{''}$ at the
detected threshold 1 mJy.

We have tried to identify the pulsars from the FIRST catalog, and then to
investigate their proper motions.

\section{IDENTIFICATIONS}

Only 42 known pulsars listed in PSR catalog are located in the sky region
covered by the FIRST. We searched for radio sources in 
the FIRST catalog within $30^{''}$ around each 
pulsar, and 17 radio sources have been found. At the 1 mJy threshold, FIRST
detected $\sim$ 90 sources per square degree.
The probability for chance coincidence in
the region of $30^{''}$ angular radius is about 0.6\%. We noticed that
pulsar positions in PSR catalog and in FIRST survey are at different epochs.
If a pulsar has a large proper
motion, e.g. 400 mas per year (the largest known), then after 30 years, the position offset would
be $12^{''}$. So, $30^{''}$ area should not miss any known pulsar
if it is detectable by the FIRST.
 
Position of pulsars from PSR catalog generally have a typical
positional uncertainty better than $0.1^{''}$, but occasionally up to a few
arcsec for newly discovered pulsars. The 17 sources are listed in Table 1
with PSR Jnames in column 1, 
their positions in column 2 and column 3 with uncertainties in
brackets, the epochs for the position and flux densities at 1.4 GHz from PSR
catalog in column 4 and column 5, respectly. For comparison, we also list
their parameters from the FIRST	survey in Table 1: the positions in
column 6 and 7, the positional
uncetainties in column 8, the flux densities in column 10. 

The sky region of FIRST survey was done over a long period,
from 1994 to 1999 and had no accurate epoch listed in the FIRST catalog, so
we estimated the approximate epoch year by the color displayed in the images
given in the FIRST observations, and  
took MJD in the middle of each year as the approximate observation
epoch (column 9 in Table 1). This should be fine for proper motion discussions considering the position uncertainties of the FIRST
sources listed in Table1.

13 pulsars were identified (in Table 1 except J1115+5030) by comparing the peak flux densities of sources between
the FIRST and PSR catalog. The consistency of the pulsar flux
densities between the two catalogs is showed in Figure 1.
\begin{figure}[!ht]
\centering
\includegraphics[width=3in,angle=-90]{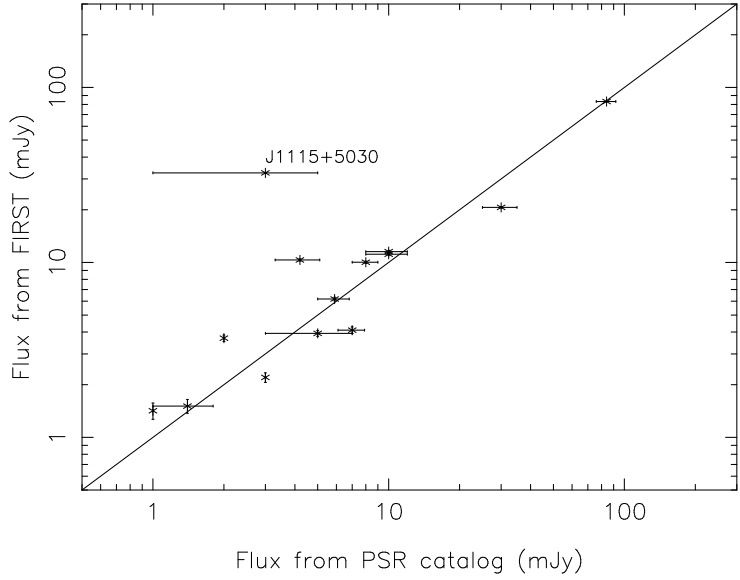}
\caption{Comparison of pulsar flux densities from the PSR catalog and the FIRST measurement}\label{figure:graph}
\end{figure}

\section{DISCUSSION}
\subsection{Undetected Pulsars}
VLA measurements of the flux densities at 1.4 GHz of most identified
pulsars are comparable to the flux densities listed in PSR catalog. PSR
J1115+5030 has a 
flux density of 
32.50 mJy in FIRST but 3 mJy in PSR catalog (Fig. 1), and its position
offset is 
$12^{''}.17$. If so, we obtained its proper motion
 $\mu_{\alpha}cos\delta$=$-$73$\pm$24 mas yr$^{-1}$,
$\mu_{\delta}$=691$\pm$24 mas yr$^{-1}$, which are much larger than those
listed in PSR catalog, $\mu_{\alpha}cos\delta$=$-$22$\pm$3 mas yr$^{-1}$,
$\mu_{\delta}$=$-$51$\pm$3 mas yr$^{-1}$.
So, we think the FIRST source is probably chance coincident, and the pulsar
was not detected in the survey.

Most of the undetected pulsars may be either lower than survey threshold 1 mJy
or influenced by the interstellar scintillation (e.g. Gupta et al. 1994)
which both helps and hinders the detections (Cordes \& Lazio 1991). In table
2, we listed those pulsars with flux densities larger than the threshold 1
mJy but did not detected in FIRST survey. 

\subsection{Flux Densities, Positions and Proper Motions}
We can see from Table 1 that PSR J1518+4904, J1640+2224, J1652+2651 have no
flux densities listed in PSR catalog. We believe the flux densities and the
positions in FIRST are more reliable than those in PSR catalog. Much
accurate positions for  three pulsars J1022+1001, J1518+4904 and J1652+2651
(see Table 1) can be used. 

The position derived from the FIRST observations were compared with those of pulsars in PSR catalog and the proper
 motions were calculated if possible. The results are listed in Table
 3. Proper motions in right ascension and declination direction  
 available in the PSR catalog are also listed in column 2 and 3 for
 comparison. Columns 4 and 5 gave the position offsets. Proper motions or
 upper limits we
 obtained are listed in columns 6 and 7. We obtained the upper limits of proper
 motions for other 3 pulsars PSRs J0751+1807, J1012+5307, J1640+2224.
 The proper motions of 10 pulsars are consistent well with those listed in PSR
 catalog (Fig. 2).
\begin{figure}[!ht]
\centering
\begin{minipage}[t]{\textwidth}
\includegraphics[width=5cm,totalheight=11cm,angle=-90]{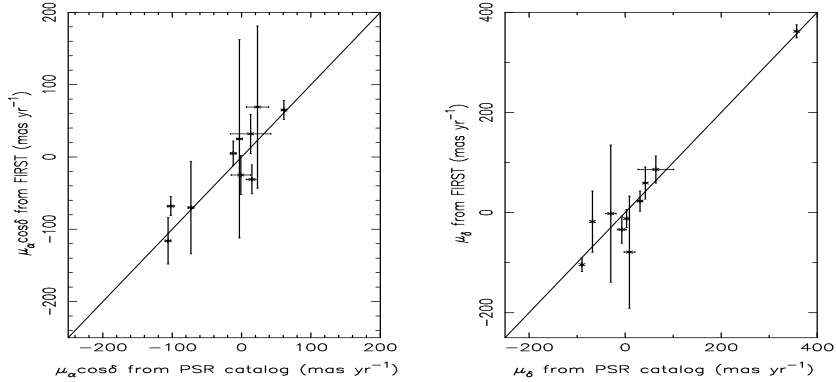}
\caption{Comparison of pulsar proper motions in ascension (left) and
declination (right)
directions listed in the PSR catalog and derived from FIRST}\label{figure:graph}
\end{minipage}
\end{figure}
\section{summary}
 In the sky region covered by the
FIRST, 17 sources near pulsar positions have been found. In the following, we
compared the pulsar positions and flux densities from the FIRST with those from
the PSR catalog (updated version of Taylor et al. 1993). We obtained new flux densities of 6 pulsars and much
accurate positions of 3 pulsars PSRs J1022+1001, JJ1518+4904, J1652+2651. For 10 pulsars, we obtained their proper
motion values consistent with those in PSR catalog, and for other 3 pulsars,
proper motion upper limits were obtained.
\acknowledgements
We thank Sun Xiaohui and Lu Yu for helpful discussions.
\oddsidemargin -10mm
\begin{table}[htbp]
\tabcolsep 1mm
\caption{Pulsar positions and flux densities from the PSR catalog and the values
of nearby FIRST sources within $30^{''}$}\label{tab:}
\begin{tabular}{cllllllllr@{$\pm$}l}
\noalign{\smallskip} 
\hline
\hline
\noalign{\smallskip} 
PSR   & RA (2000) & DEC (2000)& Epoch&Flux &  RA (first) & DEC (first) &$\sigma$&Epoch&\multicolumn{2}{c}{Flux}  \\
Jname &$\;\,h\;m\;\,s$&$\;\;\;\;\,\circ\;\;\;\;'\;\;\;\;''$& MJD& mJy&$\;\,h\;m\;\,s$&$\;\;\;\;\,\circ\;\;\;\;'\;\;\;\;''$& &MJD&\multicolumn{2}{c}{mJy}\\
\hline 					             							
\noalign{\smallskip} 			             							
0751+1807&  07 51 09.1582(7) &  +18 07 38.71(5)& 49301 &1     &07 51 09.148&+18 07 38.72&0.89&50996&  1.42 &0.151\\
0826+2637&  08 26 51.310(2) &  +26 37 25.57(7)& 40264 &10(2)&08 26 51.438& +26 37 22.83&0.35&49901&  11.14 &0.135\\
0922+0638&  09 22 13.977(3) &  +06 38 21.69(4)& 46573 &4.2(9)&09 22 14.005&+06 38 22.82&0.36&51361&  10.33& 0.139\\
0943+1631&  09 43 30.042(4) &  +16 31 35.49(6)& 47555 &1.4(4)&09 43 30.092&+16 31 34.67&1.17&51361&   1.51& 0.137\\
0953+0755&  09 53 09.316(3) &  +07 55 35.60(4)& 46058 &84(8)&09 53 09.286 &+07 55 35.94&0.28&51361&  83.13 &0.153\\	  
1012+5307&  10 12 33.4326(4) &  +53 07 02.66(1)& 49220 &3 &	10 12 33.387&+53 07 02.09&0.63&50631&   2.20& 0.138\\
1022+1001&  10 22 58.06(6) &  +10 01 54(3)& 49780 &2 &\it{10 22 58.015}&\it{+10 01 52.84}&0.54&51361&   3.69& 0.153\\
1115+5030&  11 15 38.35(2)&  +50 30 13.6(3)& 44240 &3(2)&	11 15 38.483&+50 30 25.70&0.30&50631&  32.50& 0.142\\
1136+1551&  11 36 03.296(4) &  +15 51 00.7(1)& 42364 &30(5)&11 36 03.180 &+15 51 09.62&0.31&51361&  20.62& 0.136\\
1239+2453&  12 39 40.475(3) &  +24 53 49.25(3)& 46058 &10(2)&12 39 40.386 &+24 53 49.87&0.34&49901&  11.53& 0.145\\
1509+5531&  15 09 25.724(9) &  +55 31 33.01(8)& 48383&8.0(10)&15 09 25.674&+55 31 32.90&0.37&50631&  10.02& 0.146\\
1518+4904&  15 18 16.6(1) &  +49 04 35(1)& 49896 &      &\it{15 18 16.832}&\it{+49 04 34.19}&0.45&50631&   5.03& 0.127\\
1543+0929&  15 43 38.826(6) &  +09 29 16.8(2)& 42304 &5.9(9)&15 43 38.835&+09 29 16.50&0.41&51361&   6.18& 0.146\\
1607$-$0032&  16 07 12.117(2)&$-$00 32 40.18(6)& 42307 &5(2)&16 07 12.078&$-$00 32 40.98&0.63&50996&    3.93& 0.148\\
1640+2224&  16 40 16.7417(1) &  +22 24 09.015(3)& 49360 &   &	16 40 16.698&+22 24 08.98&0.87&50996&   1.92& 0.150\\
1652+2651&  16 52 03.0(3) &  +26 51 40(1)& 49800 &   &\it{16 52 03.080}&\it{+26 51 39.85}&0.39&49901&   6.27& 0.143    \\
2145$-$0750&  21 45 50.4693(2)&$-$07 50 18.34(1)&48979&7.0(9)&21 45 50.477&$-$07 50 18.35&0.62&50631&   4.10& 0.175     \\
\noalign{\smallskip} 
\hline
\hline
\noalign{\smallskip} 
\end{tabular}
\end{table}
\newpage
\begin{table}[htbp]
\caption{Pulsars which should be detected by FIRST but not}\label{tab:}
\begin{tabular}{lllr}
\noalign{\smallskip} 
\hline
\hline
\noalign{\smallskip} 
PSR   &\multicolumn{1}{c}{RA}  &\multicolumn{1}{c}{DEC}
&\multicolumn{1}{c}{1.4 GHz}\\ 
Jname   &$\;\,h\;m\;\,s$&$\;\;\;\;\,\circ\;\;\;\;'\;\;\;\;''$&\multicolumn{1}{c}{mJy}\\
\hline 					             							
\noalign{\smallskip} 			             						
0823+0159  &	 08:23:09.76(1)&	 +01:59:12.8(5)&1.5(7)\\	      
0837+0610  &	 08:37:05.649(3)&	 +06:10:14.08(5)& 4.0(10)\\	      
\noalign{\smallskip} 
\hline
\hline
\noalign{\smallskip} 
\end{tabular}
\end{table}
\begin{table}[!htbp]
\caption{Pulsar proper motions or upper limits}\label{tab:}
\begin{tabular}{cr@{$\pm$}lr@{$\pm$}lr@{$\pm$}lr@{$\pm$}lr@{$\,\pm\,$}lr@{$\,\pm\,$}ll}
\noalign{\smallskip} 
\hline
\hline
\noalign{\smallskip} 
PSR              &\multicolumn{2}{c}{$\mu_{\alpha}cos\delta$}&\multicolumn{2}{c}{$\mu_{\delta}$} &\multicolumn{2}{c}{$\Delta$RA}&\multicolumn{2}{c}{$\Delta$DEC}&\multicolumn{2}{c}{$ \mu_{\alpha}cos\delta$}&\multicolumn{2}{c}{$\mu_{\delta}$}&Notes    \\ 
Jname   &\multicolumn{2}{c}{mas yr$^{-1}$} &\multicolumn{2}{c}{mas yr$^{-1}$}&\multicolumn{2}{c}{$''$}&\multicolumn{2}{c}{$''$}&\multicolumn{2}{c}{ mas yr$^{-1}$}  &\multicolumn{2}{c}{mas yr$^{-1}$}& \\ 
\hline 					             							
\noalign{\smallskip} 			             						
0751+1807&\multicolumn{2}{c}{...}&\multicolumn{2}{c}{...}&$-$0.15&   0.89 &
0.01&   0.89&$-$31&   191 &     2 &  192 &pm upper limit \\
0826+2637& 61&3& $-$90&2&      1.73&   0.35 & $-$2.74&   0.36 &65&
13 &  $-$104 &   14&pm consistent\\ 
0922+0638&   13&9&  64&37 &   0.42&   0.36 &  1.13&     0.36 &
32&    27 &    86 &   27&pm consistent \\ 
0943+1631& 23&16&   9&11 &     0.72&   1.17 & $-$0.82&   1.17 & 69&
112 &   $-$79 &  112&pm consistent \\ 
0953+0755& 15&8&  31& 5 &     $-$0.45&   0.29 &  0.34&   0.29 &
$-$31&    20 &    23 &   20&pm consistent\\ 
1012+5307&\multicolumn{2}{c}{...}&\multicolumn{2}{c}{...}& $-$0.41&   0.63 & $-$0.57& 0.63 &
$-$106&   163 &  $-$147 &  163&pm upper limits\\ 
1022+1001&\multicolumn{2}{c}{...}&\multicolumn{2}{c}{...}&$-$0.66&   1.05 & $-$1.16& 3.05 &
\multicolumn{2}{c}{...} &\multicolumn{2}{c}{...}&new position\\ 
1115+5030& 22&3& $-$51& 3 &   $-$1.28&   0.42 & 12.10&   0.42 &
\multicolumn{2}{c}{...}&\multicolumn{2}{c}{...}&no detection \\ 
1136+1551& $-$102&5& 357&3 & $-$1.68&   0.32 &  8.92&   0.33 &
\bf{$-$68}&\bf{13} &   362 &   13&pm consistent\\ 
1239+2453& $-$106&4&  42&3 & $-$1.22&   0.34 &  0.62&   0.34 &
$-$116&    32 &    59 &   32& pm consistent \\ 
1509+5531& $-$73&4&$-$68& 3&  $-$0.43&   0.39 & $-$0.11& 0.38 &
$-$70&    64 &   $-$18 &   61&pm consistent \\ 
1518+4904&\multicolumn{2}{c}{...}&\multicolumn{2}{c}{...} &  2.28&   1.57 & $-$0.81&   1.10 &
\multicolumn{2}{c}{...} &\multicolumn{2}{c}{...}&new position \\ 
1543+0929& $-$12&4& 3& 3 &    0.13&   0.42 & $-$0.30&   0.45 &   5&
17 &   $-$12 &   18&pm consistent \\ 
1607$-$0032&$-$1.0&14&$-$7&9&   $-$0.58&   0.63 &$-$0.80&0.63 &
$-$25&    27 &   $-$34 &   27&pm consistent \\ 
1640+2224&\multicolumn{2}{c}{...}&\multicolumn{2}{c}{...}& $-$0.61&   0.87 & $-$0.04& 0.87 &
$-$136&   193 &    $-$8 &  193&pm upper limits \\ 
1652+2651&\multicolumn{2}{c}{...}&\multicolumn{2}{c}{...}&  1.08&   4.52 & $-$0.15&   1.07 &
\multicolumn{2}{c}{...} &\multicolumn{2}{c}{...}&new position \\ 
2145$-$0750&$-$3&4&$-$30&11&     0.11&   0.62 & $-$0.01& 0.62 & 25&
137 &    $-$2 &  137&pm consistent \\ 
\noalign{\smallskip} 
\hline
\hline
\noalign{\smallskip} 
\end{tabular}
\end{table}


\newpage
\begin{thebibliography}{99}
\bibitem{}Becker R. H., White R. L., Helfend D. J. 1995, ApJ, 450, 559
\bibitem{}Bell J. F., Bailes M., Manchester R. N., Weisberg J. M., Lyne
A. G., 1995, ApJ, 440, L81
\bibitem{}Cordes J. M., Lazio T. J., 1991, ApJ, 376, 123
\bibitem{}Gupta Y., Richett B. J., Lyne A. G., 1994, MNRAS, 269, 1035
\bibitem{}Kaspi V. M., Taylor J. H., Ryba M.F., 1994, ApJ, 428, 713
\bibitem{}McGary R. S., Brisken W. F., Fruchter A. S., Goss W. M.,
Thorsett S. E., 2001, AJ, 121, 1192 
\bibitem{}Mignani R. P., De Luca A., Caraveo P. A. 2000, ApJ, 543, 318
\bibitem{}Nice D. J., Taylor J. H., 1995, ApJ, 441, 429
\bibitem{}Taylor J. H., Manchester R. N., Lyne A. G., 1993, APJS, 88, 529
\bibitem{}Wolszczan A., Doroshenko O., Konacki M., Kramer M., Jessner
A., Weilebinski R., Camilo F., Nice D. J., Taylor J. H., 2000, ApJ, 528, 907
\end{thebibliography}
\end{document}